\documentclass[twocolumn,english]{revtex4}
\usepackage[T1]{fontenc}
\usepackage[utf8]{luainputenc}
\usepackage{amsmath}
\usepackage{amssymb}
\usepackage{graphicx}
\usepackage{esint}

\makeatletter
\@ifundefined{textcolor}{}
{%
 \definecolor{BLACK}{gray}{0}
 \definecolor{WHITE}{gray}{1}
 \definecolor{RED}{rgb}{1,0,0}
 \definecolor{GREEN}{rgb}{0,1,0}
 \definecolor{BLUE}{rgb}{0,0,1}
 \definecolor{CYAN}{cmyk}{1,0,0,0}
 \definecolor{MAGENTA}{cmyk}{0,1,0,0}
 \definecolor{YELLOW}{cmyk}{0,0,1,0}
}

\usepackage{babel}

\usepackage{babel}

\usepackage{babel}

\usepackage{babel}

\usepackage{babel}

\usepackage{babel}

\usepackage{babel}

\usepackage{babel}

\usepackage{babel}

\usepackage{babel}

\usepackage{babel}

\usepackage{babel}

\usepackage{babel}

\usepackage{babel}

\usepackage{babel}

\usepackage{babel}

\usepackage{babel}

\usepackage{babel}

\usepackage{babel}

\usepackage{babel}

\usepackage{babel}

\usepackage{babel}

\makeatother

\usepackage{babel}
\begin{document}

\title{Equilibration and GGE for hard wall boundary conditions}

\author{Garry Goldstein and Natan Andrei}

\affiliation{Department of Physics, Rutgers University, Piscataway, New Jersey
08854, USA}
\begin{abstract}
In this work we present an analysis of a quench for the repulsive
Lieb-Liniger gas confined to a large box with hard wall boundary conditions.
We study the time average of local correlation functions and show
that both the quench action approach and the GGE formalism are applicable
for the long time average of local correlation functions. We find
that the time average of the system corresponds to an eigenstate of
the Lieb-Liniger Hamiltonian and that this eigenstate is related to
an eigenstate of a Lieb-Liniger Hamiltonian with periodic boundary
conditions on an interval of twice the length and with twice as many
particles (a doubled system). We further show that local operators
with support far away from the boundaries of the hard wall have the
same expectation values with respect to this eigenstate as corresponding
operators for the doubled system. We present an example of a quench
where the gas is initially confined in several moving traps and then
released into a bigger container, an approximate description of the
Newton cradle experiment. We calculate the time average of various
correlation functions for long times after the quench. 
\end{abstract}
\maketitle
\label{sec:Introduction}\textit{Introduction.} Nonequilibrium many
body physics is one of the most challenging areas of research of modern
condensed matter physics. There have been spectacular advances in
the field, driven by experimental studies of dynamics in optically
trapped atomic gas systems, systems with extremely weak coupling to
the environment allowing a study of essentially Hamiltonian dynamics
of time evolution \cite{key-5-1,key-6,key-1-1,key-3-1,key-7,key-8,key-9-1,key-11,key-4-1}.
Encouraged by these experimental advances there has been great theoretical
activity in the area \cite{key-12,key-13,key-14,key-15,key-16,key-17,key-18,key-19,key-8-1,key-25,key-26},
focused on questions like does a steady state emerge, how do local
observables equilibrate, is there any principle which allows us to
relate the steady state to the initial conditions?

One of the most important recent experimental \cite{key-5-1,key-55}
and theoretical \cite{key-10,key-34,key-33,key-35,key-36,key-37,key-38,key-39,key-40,key-41,key-42,key-3,key-32}
results is that there is a relation between the initial state and
the long time steady state for time evolution of integrable models.
This was ascribed to the fact that integrable models possess an infinite
family of local conserved charges $\left\{ I_{i}\right\} $, in involution,
which include the Hamiltonian $H$, typically identified with $I_{2}$:
\begin{equation}
\left[H,I_{i}\right]=\left[I_{i},I_{i'}\right]=0,\, H=I_{2},\label{eq:Conserved_Charges-1}
\end{equation}
These conserved quantities imply that there is a complete set of eigenstates
for an integrable model which may be parametrized by sets of rapidities
$\left\{ k_{i}\right\} $ which are simultaneous eigenstates of all
$I_{i}$. For the Lieb-Liniger Hamiltonian, the model which describes
the Newton Cradle experiment \cite{key-5-1}, the the action of the
charges on these eigenstates given by: $I_{i}\left|\left\{ k\right\} \right\rangle =\sum k^{i}\left|\left\{ k\right\} \right\rangle $.
It was shown for the Lieb-Liniger gas, by following its actual time
evolution numerically and analytically, \cite{key-10}, that at long
times the gas reaches equilibration with the density matrix having
no time dependence and becoming diagonal in the basis $\left\{ k_{i}\right\} $.

How to describe this diagonal, time independent, density matrix in
general is an open question. It was proposed that the diagonal ensemble
in turn \cite{key-10,key-34} takes the form of a generalized Gibbs
ensemble (GGE) \cite{key-31,key-32,key-33,key-34,key-35,key-36,key-37,key-38,key-39,key-40,key-41,key-42,key-8-2},
\begin{equation}
\rho_{GGE}=\frac{1}{Z}\exp\left(-\sum\alpha_{i}I_{i}\right)\label{eq:GGE_density_matrix-1}
\end{equation}
with the $\alpha_{i}$, the generalized inverse temperatures, encoding
the initial state $\left|\Phi_{0}\right\rangle $ through the requirement
$\langle I_{i}\rangle_{final}\equiv Tr\left\{ \rho_{GGE}I_{i}\right\} =\left\langle \Phi_{0}\left|I_{i}\right|\Phi_{0}\right\rangle \equiv\langle I_{i}\rangle_{initial}$.
$Z$ is a normalization constant insuring $Tr\left[\rho_{GGE}\right]=1$.
This interesting proposal, while valid for the case at hand, the repulsive
Lieb-Liniger model, fails for models with bound states (string solutions)
\cite{key-46}, a large class of models which encompasses, among others,
the attractive Lieb-Liniger, the XXZ Heisenberg chain and the Hubbard
model.

When a GGE description is valid it provides an elegant shortcut to
the computation of correlation functions at long times, without having
to explicitly follow the time evolution or to compute overlaps. Instead,
having reached equilibration the correlation functions (of the Lieb-Liniger
gas) at long times, or in this case the time average of the correlation
functions, may be computed by taking their expectation value with
respect to the GGE density matrix, e.g. $\left\langle \Theta\left(t\rightarrow\infty\right)\right\rangle =Tr\left[\rho_{GGE}\Theta\right]$.
It was further shown \cite{key-1} that the GGE ensemble is equivalent
to an eigenstate, $\rho_{GGE}\cong\left|\left\{ k_{0}\right\} \right\rangle \left\langle \left\{ k_{0}\right\} \right|$,
for an appropriately chosen $\left|\left\{ k_{0}\right\} \right\rangle $
so that $\left\langle \Theta\left(t\rightarrow\infty\right)\right\rangle =\left\langle \left\{ k_{0}\right\} \right|\Theta\left|\left\{ k_{0}\right\} \right\rangle $.
Another approach, the quench action approach \cite{key-52}, is of
more general validity but is more difficult to implement. It allows
the computation of the diagonal density matrix in terms of the overlaps
of eigenstates with the initial state, but such overlaps are hard
to determine and are known only for few initial states. Again, it
was shown that the resulting diagonal ensemble is equivalent to an
eigenstate.

Most of the work done on the Lieb-Liniger model was concerned with
periodic boundary conditions, exceptions are \cite{key-50}. Real
systems \cite{key-5-1,key-6,key-1-1,key-3-1,key-7,key-8,key-9-1,key-11,key-4-1}
have finite extent with typically a parabolic potential confining
the particles. We will approximate this parabolic confining potential
as a hard wall boundary. We will study the system in the limit where
the system size $L\rightarrow\infty$, the number of particles $N$
scales with the system size, $N/L=const$, and for times much greater
then the system size, $t>L/v_{typ}$ ($v_{typ}$ is a typical velocity).
This regime is relevant for many experiments. We note that this time
scale has been theoretically considered for the Tonks gas before \cite{key-49,key-50}.

\textit{Time average.} We shall consider circumstances where the system
does not necessarily equilibrate in the long time limit and focus
instead on the long time average of a local operator (observable)
$\Theta$ evolving from the initial state $\left|\Psi\right\rangle $,
\begin{equation}
\begin{array}[t]{l}
\left\langle \Theta\right\rangle _{T}\equiv\frac{1}{T}\int_{0}^{T}dt\left\langle \Psi\right|e^{iH_{LL}t}\Theta e^{-iH_{LL}t}\left|\Psi\right\rangle =\\
=\frac{1}{T}\sum_{\lambda}\sum_{\kappa}\frac{e^{i\left(E_{\lambda}-E_{\kappa}\right)T}-1}{i\left(E_{\lambda}-E_{\kappa}\right)}\left\langle \Psi\mid\lambda\right\rangle \left\langle \lambda\right|\Theta\left|\kappa\right\rangle \left\langle \kappa\mid\Psi\right\rangle \\
\cong\sum_{\lambda}\left\langle \Psi\mid\lambda\right\rangle \left\langle \lambda\right|\Theta\left|\lambda\right\rangle \left\langle \lambda\mid\Psi\right\rangle ,
\end{array}\label{eq:Diagonal_ensemble}
\end{equation}
we find it is given by a diagonal ensemble in the limit $T\rightarrow\infty$.
Therefore the time averaged expectation values of local observables
is given by the diagonal ensemble. Here $\left|\lambda\right\rangle $
and $\left|\kappa\right\rangle $ are exact eigenstates of the system
in the box.

\textit{The system} we shall study is the Lieb-Liniger Hamiltonian
describing the 1-D system of bosons with short range interactions
\cite{key-31,key-43,key-55-1}: 
\begin{equation}
H_{LL}=\intop_{0}^{L}dx\left\{ \partial_{x}b^{\dagger}\left(x\right)\partial_{x}b\left(x\right)+c\left(b^{\dagger}\left(x\right)b\left(x\right)\right)^{2}\right\} .\label{eq:lieb-lin-hamiltonian-1}
\end{equation}
Here $b^{\dagger}\left(x\right)$ is the bosonic creation operator
at the point $x$ and $c$ is the coupling constant. Hard wall boundary
conditions are imposed: 
\begin{align}
\psi\left(x_{1}=0,x_{2},...x_{N}\right) & =0\nonumber \\
\psi\left(x_{1},x_{2},..x_{N}=L\right) & =0\label{eq:Boundary_conditions}
\end{align}
with $\psi\left(x_{1},...x_{N}\right)$ the wave function of the bosons
in the region $x_{1}<x_{2}<...<x_{N}$.

The exact eigenstates of the Hamiltonian with the boundary conditions
given in Eq. (\ref{eq:Boundary_conditions}) are given by \cite{key-43}:
\begin{equation}
\psi\left(\left|k_{1}\right|,..\left|k_{N}\right|\right)=\sum_{\left\{ \varepsilon\right\} }C\left\{ \varepsilon\right\} \bar{\psi}\left(\varepsilon_{1}\left|k_{1}\right|,..\varepsilon_{N}\left|k_{N}\right|\right),\label{eq:Eigenstate}
\end{equation}
where $\left\{ \varepsilon\right\} $ corresponds to the $2^{N}$
sequences $\varepsilon_{j}=\pm1$ and $C\left(\varepsilon_{1},...\varepsilon_{N}\right)=\prod\varepsilon_{j}\prod_{i<j}\left(1-\frac{ic}{\varepsilon_{i}\left|k_{i}\right|+\varepsilon_{j}\left|k_{j}\right|}\right)$,
and 
\[
\bar{\psi}\left(k_{1},...k_{N}\right)=\sum_{P}A\left(P\right)e^{i\sum k_{P_{i}}x_{i}},\quad x_{1}<x_{2}<...<x_{N}
\]
with $A\left(P\right)=\prod_{i<j}\left(1+\frac{ic}{k_{P_{i}}-k_{P_{j}}}\right)$
and the sum $\sum_{P}$ extending over $N!$ permutations. These are
the eigenstates with periodic boundary conditions. Furthermore the
rapidities $k_{i}=\varepsilon_{i}\left|k_{i}\right|$ satisfy the
Bethe ansatz equations \cite{key-43}: 
\[
k_{i}L=\pi n_{i}+\sum_{j\neq i}\left(\arctan\left(\frac{c}{k_{i}-k_{j}}\right)+\arctan\left(\frac{c}{k_{i}+k_{j}}\right)\right)
\]
These are exactly the same equations as for a doubled system of length
$2L$ with twice as many particles having $2N$ with rapidities $\left\{ \varepsilon_{i}\left|k_{i}\right|\right\} $.
There is a one to one correspondence between states of a system with
hard wall boundary conditions and states of a doubled system with
periodic boundary conditions where all the rapidities come in pairs
$\left\{ k,-k\right\} $ \cite{key-43}. The Bethe Ansatz equations
which determine the allowed rapidities $\{k\}$ for the doubled system
can be translated in a standard fashion \cite{key-31} into a set
of integral equations for the rapidities' densities. We denote, for
a given eigenstate $\left|\left\{ k\right\} \right\rangle $ of the
doubled system, by $\rho_{p}\left(k\right)$ the Bethe density of
particles so that $2L\rho_{p}\left(k\right)dk$ is the number of particles
in the interval $\left[k,k+dk\right]$ of the doubled system. Similarly
$\rho_{h}\left(k\right)$ denotes the hole density and $\rho_{t}\left(k\right)=\rho_{p}\left(k\right)+\rho_{h}\left(k\right)$
the total density. The number of states $\left|\left\{ k\right\} \right\rangle $,
consistent with a given set of densities, $\{\rho_{p}\left(k\right),\rho_{h}\left(k\right)\}$,
is measured by the Yang-Yang entropy \cite{key-31}, $S_{YY}\left(\left\{ \rho\right\} \right)=\int_{-\infty}^{\infty}dk\left(\rho_{h}\left(k\right)\ln\left(\frac{\rho_{t}\left(k\right)}{\rho_{h}\left(k\right)}\right)+\rho_{p}\left(k\right)\ln\left(\frac{\rho_{t}\left(k\right)}{\rho_{p}\left(k\right)}\right)\right)$.
The densities $\{\rho_{p}\left(k\right),\rho_{h}\left(k\right)\}$
for the doubled system are determined by the thermodynamic Bethe Ansatz
equations which enforce the periodic boundary conditions: $\rho_{t}\left(k\right)=\frac{1}{2\pi}+\frac{1}{2\pi}\int K\left(k,q\right)\rho_{p}\left(q\right)$,
with $K\left(k,q\right)=\frac{2c}{c^{2}+\left(k-q\right)^{2}}$.

\textit{Time average, quench action and GGE action.} The time average
of a local observable, Eq.(\ref{eq:Diagonal_ensemble}), can be rewritten
as \cite{key-44,key-52}: 
\begin{equation}
\left\langle \Theta\right\rangle _{T\rightarrow\infty}=\int D\left(\frac{\rho_{t}}{\rho_{p}}\right)e^{2LS_{Quench}\left(\left\{ \rho\left(k\right)\right\} \right)}\left\langle \left\{ k\right\} \right|\Theta\left|\left\{ k\right\} \right\rangle \label{eq:Quench_integral}
\end{equation}
Here $\left\langle \left\{ k\right\} \right|\Theta\left|\left\{ k\right\} \right\rangle $
is computed in the hard wall (non-doubled) system, and the quench
action is given by: 
\[
S_{Quench}\left(\left\{ \rho\left(k\right)\right\} \right)=\int g^{\Phi}\left(k\right)\rho\left(k\right)+\frac{1}{2}S_{YY}\left(\left\{ \rho\left(k\right)\right\} \right)
\]
with $\int g^{\Phi}\left(k\right)\rho\left(k\right)=\frac{2}{2L}\log\left(\left|\left\langle \Phi\mid\left\{ k\right\} \right\rangle \right|\right)$,
where $g\left(k\right)=g\left(-k\right)$. The extra factor of $\frac{1}{2}$
in front of $S_{YY}\left(\left\{ \rho\left(k\right)\right\} \right)$
comes from the fact that we are only considering states where the
rapidities come in pairs $\left\{ k,-k\right\} $.

The time average of the Lieb-Liniger gas with hard wall boundary conditions
corresponds therefore to a single eigenstate the one that maximizes
the quench action \cite{key-44}. Let us denote solution quasiparticle
density for the doubled system as $\rho_{p}^{\Phi}\left(k\right)$
and the quasiparticle density for the original system as $\tilde{\rho}_{p}^{\Phi}\left(k\right)$
with $\tilde{\rho}_{p}^{\Phi}\left(k\right)=2\theta\left(k\right)\rho_{p}^{\Phi}\left(k\right)$
and $\rho_{p}^{\Phi}\left(k\right)=\rho_{p}^{\Phi}\left(-k\right)$.

We proceed to convert the quench action into a GGE description of
the system with hard wall boundary conditions and use it to compute
time average of local observables. We begin by determining its conserved
charges. Since an eigenstate, see Eq. (\ref{eq:Eigenstate}) is given
by a superposition of rapidities of the form $\left\{ \varepsilon_{i}\left|k_{i}\right|\right\} $
all the even conserved quantities satisfy the relation: 
\begin{equation}
I_{2n}\psi\left(\left|k_{1}\right|,..\left|k_{N}\right|\right)=\sum k_{i}^{2n}\psi\left(\left|k_{1}\right|,..\left|k_{N}\right|\right)\label{eq:Conserved_quantities}
\end{equation}
Therefore $\left\{ I_{2n}\right\} $ form a set of local integrals
of motion and the quasiparticle density $\rho_{p}^{\Phi}\left(k\right)$
being symmetric in $k$ is, in turn, uniquely determined by its even
moments which correspond to its conserved quantities \cite{key-54}
in particular $\left\{ I_{2n}\right\} $ are complete. Hence the even
local integrals of motion, $\left\{ I_{2n}\right\} $, determine final
state in terms of the GGE density operator, $\tilde{\rho}_{GGE}=\frac{1}{Z}\exp\left(-\sum\alpha_{2n}I_{2n}\right)$.
The inverse temperatures, $\alpha_{2n}$, found from the initial state
$|\Phi\rangle$ setting $\left\langle I_{2n}\right\rangle =\left\langle I_{2n}\right\rangle _{GGE}$.
Any local observable can be written in the form: 
\begin{equation}
\left\langle \Theta\right\rangle _{GGE}=\int D\left(\frac{\rho_{t}}{\rho_{p}}\right)e^{2LS_{GGE}\left(\left\{ \rho\left(k\right)\right\} \right)}\left\langle \left\{ k\right\} \right|\Theta\left|\left\{ k\right\} \right\rangle \label{eq:GGE_Path-integral}
\end{equation}
with $S_{GGE}=\int\rho\left(k\right)\left(\frac{1}{2}\sum\alpha_{2n}k^{2n}+\frac{1}{2L}\ln Z\right)+\frac{1}{2}S_{YY}$.
We can identify $g\left(k\right)=\frac{1}{2}\sum\alpha_{2n}k^{2n}+\frac{1}{2L}\ln Z$
(since both the quench action and the GGE are equivalent to a single
eigenstate of the Lieb Liniger Hamiltonian (which corresponds to the
extremum of the path integral in Eq. (\ref{eq:GGE_Path-integral})),
we establish that $\left\langle \Theta\right\rangle _{T\rightarrow\infty}=\left\langle \Theta\right\rangle _{GGE}$.
We conclude that the time average of Lieb-Liniger gas corresponds
to a GGE density matrix where the conserved operators are the even
local conserved densities. We further note that when considering an
operator $\Theta$ with support far away from from the hard wall boundaries
we may as well calculate $\left\langle \Theta\right\rangle _{GGE}$
with respect to the doubled system. Indeed $\left\langle \Theta\right\rangle _{GGE}=Tr\left\{ \Theta\exp\left(-\sum\alpha_{2n}I_{2n}\right)\right\} $
for both systems. Since operators $I_{2n}$ are local, all correlation
functions $\Theta$ may be calculated by considering paths where the
propagator is the quadratic piece of $\sum\alpha_{2n}I_{2n}$ while
the interactions are given by the quartic and higher order pieces.
We note that paths that cross the boundary of the system are exponentially
suppressed when $\Theta$ is far from the boundary. 
\begin{figure}
\begin{centering}
\includegraphics[bb=0bp 150bp 720bp 350bp,width=1\columnwidth]{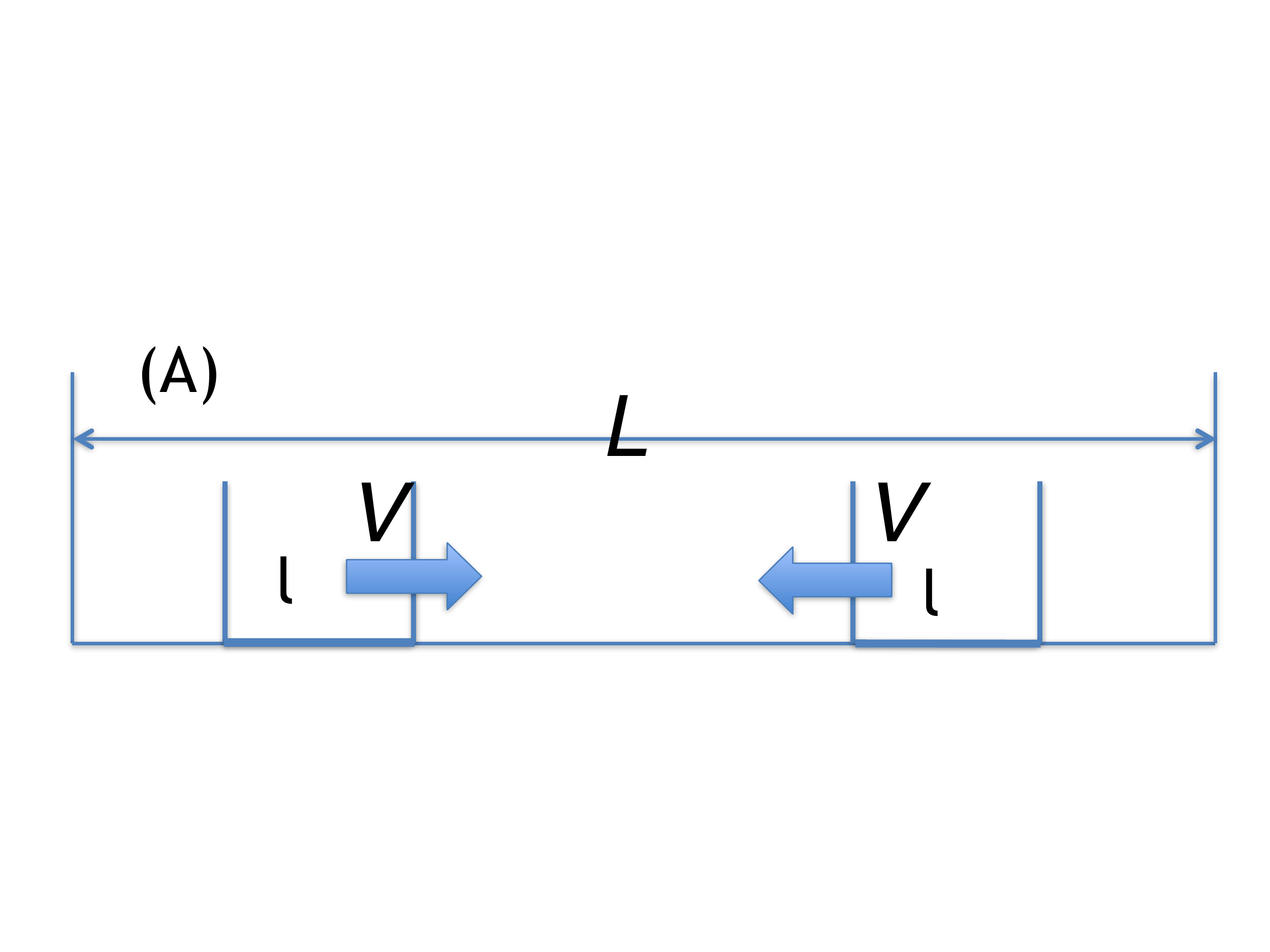}
\includegraphics[scale=0.25]{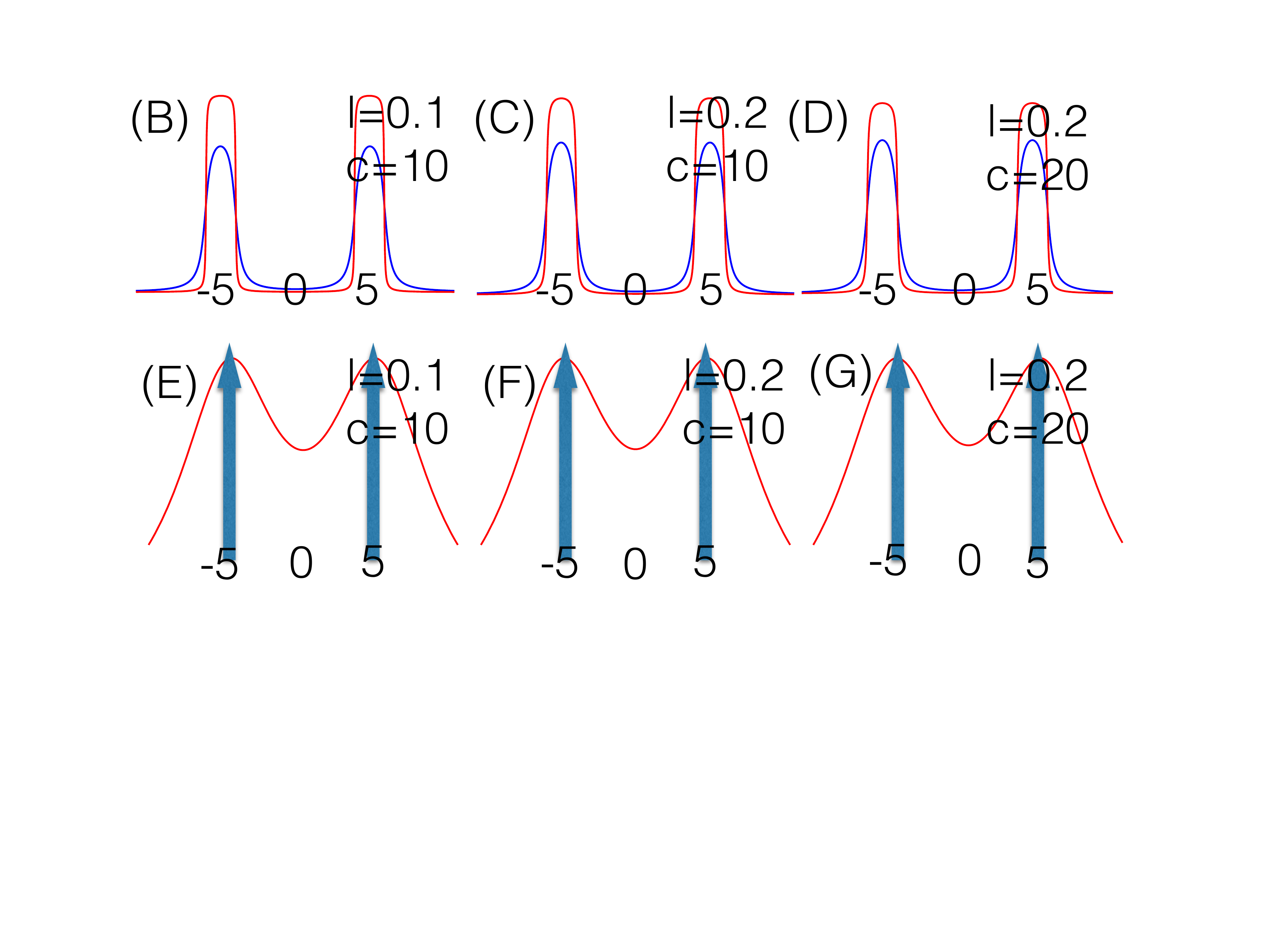} \vspace{-3cm}

\par\end{centering}

\protect\protect\caption{\label{fig:Poles} (A) The system is initialized in a state where
two hard wall Lieb Liniger droplets of length $L_{i}$ moving with
velocity $\pm V$ inside a large hard wall trap of length $L$. (B-G)
The velocity distribution for the BEC bottom and the ground state
quench top for a variety of quasiparticle densities and interaction
strengths. The time average velocity distribution in red the initial
velocity distribution before the quench is shown in blue. (B-D) $V=5$,
$L=1$, $k_{F}=1$. (E-G) $V=5$, $L=1$ $n=1$. The initial velocity
distribution is computed in the supplementary online information and
is shown in blue while the final velocity distribution is shown in
red. the initial velocity distribution of the BEC is shown in the
form of delta functions. Note collision narrowing in (B-D) and broadening
in (E-G).}
\end{figure}

\textit{Examples: 1. Newton's cradle type - eigenstate initial conditions.
} We will consider the following setup: there is a large trap of length
$L$ with hard wall boundary conditions in which there are multiple
smaller traps of lengths $L_{i}$ moving with velocities $V_{i}$.
Each of the smaller traps contains a Lieb-Liniger gas initialized
in an eigenstate described by quasiparticle density $\rho_{p}^{i}\left(k\right)$
with $\rho_{p}^{i}\left(k\right)=\rho_{p}^{i}\left(-k\right)$ (we
note that thermal states also correspond to specific eigenstates \cite{key-31}).
At time $t=0$ the smaller traps are turned off and the whole of the
gas expands into the larger trap. We would like to find the quasiparticle
density of the long time averaged final state. To do so we use the
fact that all the even local conserved quantities are conserved during
the quench, so we need to equate their values before and after the
quench. We will show in the supplementary online information that
in the thermodynamic limit we do not need to consider the edge effects
for computing the local conserved quantities. Therefore we need to
find a symmetric quasiparticle density that satisfies the following
set of equations: 
\[
L\int dk\rho_{p}^{f}\left(k\right)k^{2n}=\sum L_{i}\int\rho_{p}^{i}\left(k\right)\left(k+\frac{1}{2}V_{i}\right)^{2n}
\]
The extra terms $k+\frac{1}{2}V_{i}$ stem from the fact that under
a boost to velocity $V_{i}$ the wave function is multiplied by $\exp\left(i\sum m_{i}V_{i}x_{i}\right)$
with $m_{i}=\frac{1}{2}$. We note that here $\rho_{p}^{i}\left(k\right)=\rho_{p}^{i}\left(-k\right)$
is the quasiparticle density of the doubled system. A solution to
this equation is given by: 
\[
\rho_{p}^{f}\left(k\right)=\sum\frac{L_{i}}{2L}\left(\rho_{p}^{i}\left(k+\frac{1}{2}V_{i}\right)+\rho_{p}^{i}\left(k-\frac{1}{2}V_{i}\right)\right),
\]
This solution allows for the calculation of various correlation functions
for the system. Note that in the case of a periodic boundary condition
we would have received the answer $\rho_{p}^{f}\left(k\right)=\sum\frac{L_{i}}{L}\rho_{p}^{i}\left(k-\frac{1}{2}V_{i}\right)$.

Consider now the quench dynamics of a system consisting initially
of two boxes of length $l$ with $N$ particles each in the ground
state moving with of opposite velocities $V$ and $-V$ see Fig. \ref{fig:Poles}(A).
In experiment one typically measures the probability distribution
for the particle velocity. It is given by the Fourier transform of
the field-field correlation function 
\[
P\left(v,t\right)=\int dxe^{-i\frac{v}{2}x}\left\langle b^{\dagger}\left(x\right)b\left(0\right)\right\rangle _{t}
\]
We will be interested in its time average. This example has many similarities
to the experiment done by Kinoshita et. al \cite{key-5-1} where the
system is placed in a parabolic confining potential and initialized
in a state with some of the particles going left and some of the particles
going right. Here we have replaced the parabolic confining potential
with a hard wall box and do not therefore expect this probability
distribution to match well with the one measured in the experiment.
The reason being that when confined by a harmonic potential the bosons
move up and down the potential which slows them down and speeds them
up periodically. In our setup the particles hit a hard wall and have
their velocities reversed after the collision (as such they experience
no intermediate velocities). As a result our calculation is expected
to underestimate the probability of a particle having low velocity.

We now proceed with the calculation: An important ingredient in calculating
correlation functions is the occupation probability of the doubled
box $f_{L}\left(k\right)=\frac{\rho_{p}\left(k\right)}{\rho_{t}\left(k\right)}$.
To calculate it we first calculate the quasiparticle distributions
of the smaller boxes. The ground state total density $\rho_{t}$ of
the smaller boxes, in the limit of large $c$ is determined from \cite{key-31}:
$2\pi\rho_{t}\left(k\right)=1+\frac{2}{c}\int_{-k_{F}}^{k_{F}}dq\rho_{t}\left(q\right)$
leading to $\rho_{t}\left(k\right)=\theta\left(-k_{F},k_{F}\right)\frac{1}{2\pi}\left(1+\frac{2k_{F}}{\pi c}\right)+o(\frac{k_{F}}{c})....$.
Furthermore it is possible to obtain a relation between $k_{F}$ and
$N$ with $k_{F}=\frac{\pi N}{l}-\frac{2}{\pi c}\left(\frac{\pi N}{l}\right)^{2}+.....$
which implies that the total particle density of the doubled box is
given by: $\rho_{p}\left(k\right)=\frac{l}{2\pi L}\left(1+\frac{2k_{F}}{\pi c}\right)\sum_{v=\pm\frac{V}{2}}\theta\left(-k_{F}+v,k_{F}+v\right)$,
therefore the final total density is: $\rho_{t}\left(k\right)=\frac{1}{2\pi}\left(1+\frac{4k_{F}l}{c\pi L}\right)+....$
and occupation probability: 
\begin{align}
f_{L}\left(k\right)=A_{L}\sum_{v=\pm\frac{V}{2}}\theta\left(-k_{F}+v,k_{F}+v\right)\label{eq:Density}
\end{align}
with $A_{L}=\frac{l}{L}\left(1+\frac{2k_{F}}{\pi c}\left(1-\frac{2l}{L}\right)\right)$.
We now proceed to compute the field-field correlation function $\left\langle b^{\dagger}\left(x\right)b\left(0\right)\right\rangle $.
We will only consider the case when the points $x$ and $0$ are far
away from the boundaries of the box so we may use the doubled system
for all calculations. In terms of the occupation distribution, $f_{L}(k)$,
the correlation function is given by \cite{key-53}: $\left\langle b^{\dagger}\left(x\right)b\left(0\right)\right\rangle =\int\frac{dk}{2\pi}f_{L}\left(k\right)e^{-ikx}\omega\left(k\right)\exp\left(-x\int duf_{L}\left(t\right)P_{u}\left(k\right)\right)+h.o.t.$
with $\omega\left(k\right)=\exp\left(-\frac{1}{2\pi}\int dqK\left(k,q\right)f\left(q\right)\right)\cong\exp\left(-\frac{F_{L}}{\pi c}\right)$,
and $F_{L}=4k_{F}A_{L}$ and $K\left(k,q\right)=\frac{2c}{\left(k-q\right)^{2}+c^{2}}\cong\frac{2}{c}$.
The generating function $P_{u}\left(k\right)$ satisfies the equation:
$2\pi P_{u}\left(k\right)=-\frac{k-u+ic}{u-k+ic}\exp\left(-\int f_{L}\left(s\right)K\left(u,s\right)P_{s}\left(k\right)\right)-1$
yielding for large $c$: $P_{u}\left(k\right)\cong-\frac{1}{2\pi}\left(1+\exp\left(-\frac{2F_{L}}{\pi c}\right)\right)+i\frac{\exp\left(-\frac{2F_{L}}{\pi c}\right)}{\pi c}\left(k-u\right)$
and $\int f_{L}\left(u\right)P_{u}\left(k\right)\cong-\frac{F_{L}}{2\pi}\left(1+\exp\left(-\frac{2F_{L}}{\pi c}\right)\right)+iF_{L}\frac{k}{\pi c}\exp\left(-\frac{2F_{L}}{\pi c}\right)$.
Combing all we obtain the velocity probably distribution:

\begin{align}
P\left(v\right)\sim A_{L}\frac{\exp\left(-\frac{F_{L}}{\pi c}\right)}{2\pi}\sum_{i,j=\pm}\left(-1\right)^{j}\arctan A_{i,j}(v)\label{velProb}
\end{align}
with $A_{\pm\pm}(v)=C_{L}\left((1-F_{L}\frac{\exp\left(-2F_{L}/\pi c\right)}{\pi c})\left(\pm\frac{V}{2}\pm k_{F}\right)+\frac{v}{2}\right)$
and $C_{L}=\frac{2\pi}{4k_{F}A_{L}\left(1+\exp\left(-\frac{2F_{L}}{\pi c}\right)\right)}$.

Note that the velocity distribution Eq.(\ref{velProb}), see Fig.
\ref{fig:Poles}(B-D), underwent a collision narrowing. The distribution
is the leading order term for the set up of a hard wall trap. In a
harmonic trap, as argued before, the probability of a particle having
low velocity would be larger due to having to move up and down the
harmonic confining potential.

\textit{2. Newton's cradle type, BEC initial conditions.} A very similar
scenario happens when we initialize the state in a collection of BEC's
each of length $L_{i}$ moving with velocity $V_{i}$ inside a larger
trap of length $L$. At $t=0$ the smaller traps are released and
interactions are turned on so that the system is described by a Lieb-Liniger
Hamiltonian with coupling constant $c$. The initial state is BEC
and can be described by a quasiparticle density \cite{key-44}: 
\begin{equation}
\rho_{p}^{BEC_{i}}\left(x\right)=\frac{\tau_{i}\frac{d}{d\tau_{i}}a\left(x,\tau\right)}{1+a\left(x,\tau\right)}.\label{eq:Quasiparticle_BEC}
\end{equation}
Here $x=\frac{k}{c}$, $\tau_{i}=\frac{n_{i}}{c}$ (where $n_{i}$
is the particle density) and $a\left(x,\tau\right)=\frac{2\pi\tau}{x\sinh\left(2\pi x\right)}J_{1-2ix}\left(4\sqrt{\tau}\right)J_{1+2ix}\left(4\sqrt{\tau}\right)$.
With $J_{n}$ a modified Bessel function of the first kind of order
$n$. By an argument similar to the one given above the final quasiparticle
density is given by: 
\begin{equation}
\rho_{p}^{f}\left(k\right)=\sum\frac{L_{i}}{2L}\left(\rho_{p}^{BEC_{i}}\left(k+\frac{1}{2}V_{i}\right)+\rho_{p}^{BEC_{i}}\left(k-\frac{1}{2}V_{i}\right)\right)\label{eq:Density_solution-1}
\end{equation}

More generally any translationally invariant quench that may be solved
using periodic boundary conditions it is possible to define a box
quench which may be solved analogously to Eq. (\ref{eq:Density_solution-1})
above. In the supplementary online information we show that for a
quench with two boxes (each of length $l$ with $N$ particles in
each box in a BEC state) with velocities $V$ and $-V$ inside of
a box of total length $L$ the velocity probably distribution is given
by: 
\begin{align*}
P\left(v\right) & \sim nB_{L}\frac{\exp\left(-\frac{G_{L}}{\pi c}\right)}{\pi}\times\\
 & \times\left(\frac{H_{L}}{H_{L}^{2}+\frac{1}{4}\left(v-VK_{L}\right)}+\frac{H_{L}}{H_{L}^{2}+\frac{1}{4}\left(v+VK_{L}\right)}\right)
\end{align*}
Here, $K_{L}=\left(1-G_{L}\frac{\exp\left(-2G_{L}/\pi c\right)}{\pi c}\right)$,
$G_{L}=2nB_{L}$ $B_{L}=\frac{l}{L}\frac{1}{\frac{1}{2\pi}+\frac{2N}{\pi cL}}$,
$H_{L}=\frac{G_{L}}{2\pi}\left(1+\exp\left(-\frac{2G_{L}}{\pi c}\right)\right)+2n\left(1-G_{L}\frac{\exp\left(-2G_{L}/\pi c\right)}{\pi c}\right)$
and $n=\frac{N}{l}$. See Fig. \ref{fig:Poles}(E-G).

We note that the average velocity distribution has broadened as compared
to its value at the start of the quench, while in the previous case
ground state initial conditions the distribution underwent narrowing
due to the collisions.

\textit{Conclusions.} We have studied a quench of the Lieb-Liniger
gas on an interval with hard wall boundary conditions. We introduced
a doubled system with periodic boundary conditions that is equivalent
to the one on an interval. We have shown that the GGE formalism applies
to the computation of time averages of local operators and that the
even integrals of motion form a complete set of local conserved quantities.
We have used this approach to compute a quench where there are several
small traps inside of a larger one and the smaller traps are released.
We found that the quasiparticle density is additive. We have also
calculated the expectation values of some local operators for this
quench and in particular the time averaged velocity distribution.
In the future it would be of interest to extend this work to models
with bound states.

\textbf{Acknowledgments}: This research was supported by NSF grant
DMR 1410583 and Rutgers CMT fellowship.

\part*{Supplementary online information}

\section*{Correlation functions (zero temperature case)}

We would like to calculate various correlation functions $\left\langle b^{\dagger}\left(0\right)b^{\dagger}\left(0\right)b\left(0\right)b\left(0\right)\right\rangle $,
$\left\langle b^{\dagger}\left(0\right)b^{\dagger}\left(0\right)b^{\dagger}\left(0\right)b\left(0\right)b\left(0\right)b\left(0\right)\right\rangle $,
and $\left\langle \rho\left(x\right)\rho\left(0\right)\right\rangle $
($\left\langle b^{\dagger}\left(x\right)b\left(0\right)\right\rangle $
and the quasiparticle occupation probability $f_{L}\left(k\right)=\frac{\rho_{p}\left(k\right)}{\rho_{t}\left(k\right)}$
has already been partially done in the main text) for quenches discussed
in the main text. We will consider the case where both $x$ and $0$
are far away from the boundaries of the box. In this case the problem
becomes translationally invariant and all correlations may be calculated
using a doubled box with doubled quasiparticle density (see the discussion
below Eq. (\ref{eq:GGE_Path-integral}) in the main text). As such
we may use the results found in \cite{key-49-1,key-50-1,key-51-1,key-52-1,key-53}.
We will consider the initial conditions where the system starts with
two small boxes of length $l$ with $N$ particles each, with each
box cooled to the ground state. We will also assume that the boxes
have velocities $V$ and $-V$. To make the computations tractable
we will work only in the limit of large $c$ and only to leading order
in $\frac{1}{c}$.

We now compute the local correlation functions using this occupation
probability. We begin with the correlation function $\left\langle b^{\dagger}\left(0\right)b^{\dagger}\left(0\right)b\left(0\right)b\left(0\right)\right\rangle $.
It is given by\cite{key-49-1}:\begin{widetext} 
\begin{align}
\left\langle b^{\dagger}\left(0\right)b^{\dagger}\left(0\right)b\left(0\right)b\left(0\right)\right\rangle  & =2\int\frac{dk_{1}}{2\pi}f_{L}\left(k_{1}\right)\int\frac{dk_{2}}{2\pi}f_{L}\left(k_{2}\right)\frac{\left(k_{1}-k_{2}\right)^{2}}{\left(k_{1}-k_{2}\right)^{2}+c^{2}}+....\nonumber \\
 & \cong\frac{4}{\pi^{2}}A_{L}^{2}k_{F}^{2}\left(\frac{V}{c}\right)^{2}+....\label{eq:density_density_zero}
\end{align}

For convenience we will denote $A_{L}=\frac{l}{L}\left(1+\frac{2k_{F}}{\pi c}\left(1-\frac{2l}{L}\right)\right)$.
Where the last equality is in the limit $k_{F}\ll V\ll c$. Furthermore
it is possible to obtain the density density density correlation function
in the same limit. It is given by \cite{key-49-1}: 
\begin{align}
\left\langle b^{\dagger}\left(0\right)b^{\dagger}\left(0\right)b^{\dagger}\left(0\right)b\left(0\right)b\left(0\right)b\left(0\right)\right\rangle \cong & 6\int\frac{dk_{1}}{2\pi}\int\frac{dk_{2}}{2\pi}\int\frac{dk_{3}}{2\pi}f_{L}\left(k_{1}\right)f_{L}\left(k_{2}\right)f_{L}\left(k_{3}\right)\frac{\left(k_{1}-k_{2}\right)^{2}}{\left(k_{1}-k_{2}\right)^{2}+c^{2}}\frac{\left(k_{1}-k_{3}\right)^{2}}{\left(k_{1}-k_{3}\right)^{2}+c^{2}}\frac{\left(k_{2}-k_{3}\right)^{2}}{\left(k_{2}-k_{3}\right)^{2}+c^{2}}\nonumber \\
\cong & \frac{1}{8\pi^{3}}A_{L}^{3}\left(\frac{2k_{F}}{c}\right)^{4}\left(\frac{V}{c}\right)^{4}\label{eq:density_density_density}
\end{align}

\end{widetext}Where again the last equality is true in the limit
$k_{F}\ll V\ll c$.

We now repeat the calculation of the field-field correlation function
$\left\langle b^{\dagger}\left(x\right)b\left(0\right)\right\rangle $
(already partly given in the main text). It is given by \cite{key-53}:
\begin{align*}
\left\langle b^{\dagger}\left(x\right)b\left(0\right)\right\rangle  & \cong\int\frac{dk}{2\pi}f_{L}\left(k\right)e^{-ikx}\omega\left(k\right)\times\\
 & \times\exp\left(-x\int dtf_{L}\left(t\right)P_{t}\left(k\right)\right)
\end{align*}

Here $\omega\left(k\right)=\exp\left(-\frac{1}{2\pi}\int dqK\left(k,q\right)f\left(q\right)\right)\cong\exp\left(-\frac{F_{L}}{\pi c}\right)$,
where $F_{L}=4k_{F}A_{L}$ and $K\left(k,q\right)=\frac{2c}{\left(k-q\right)^{2}+c^{2}}\cong\frac{2}{c}$.
Furthermore the function $P_{t}\left(k\right)$ satisfied the equation:
\begin{equation}
2\pi P_{t}\left(k\right)=-\frac{k-t+ic}{t-k+ic}\exp\left(-\int f_{L}\left(s\right)K\left(t,s\right)P_{s}\left(k\right)\right)-1\label{eq:Field_Field_generating_function}
\end{equation}

Using this expression it is possible to obtain that: 
\begin{align*}
P_{t}\left(k\right)\cong & -\frac{1}{2\pi}\left(1+\exp\left(-\frac{2F_{L}}{\pi c}\right)\right)\\
 & +i\frac{\exp\left(-\frac{2F_{L}}{\pi c}\right)}{\pi c}\left(k-t\right)
\end{align*}

From this we obtain that $\int f_{L}\left(t\right)P_{t}\left(k\right)\cong-\frac{F_{L}}{2\pi}\left(1+\exp\left(-\frac{2F_{L}}{\pi c}\right)\right)+i\frac{kF_{L}}{\pi c}\exp\left(-\frac{2F_{L}}{\pi c}\right)$.
Combing we obtain that :\begin{widetext} 
\begin{align}
\left\langle b^{\dagger}\left(x\right)b\left(0\right)\right\rangle  & =\frac{\exp\left(-\frac{F_{L}}{\pi c}\right)}{2\pi}\exp\left(-\frac{F_{L}x}{2\pi}\left(1+\exp\left(-\frac{2F_{L}}{\pi c}\right)\right)\right)\int f_{L}\left(k\right)e^{-ikx\left(1-F_{L}\frac{\exp\left(-2F_{L}/\pi c\right)}{\pi c}\right)}\nonumber \\
 & =\frac{2\exp\left(-\frac{F_{L}}{\pi c}\right)}{\pi x\left(1-\frac{\exp\left(-2F_{L}/\pi c\right)}{\pi c}\right)}\exp\left(-\frac{F_{L}x}{2\pi}\left(1+\exp\left(-\frac{2F_{L}}{\pi c}\right)\right)\right)A_{L}\times\nonumber \\
 & \times\sin\left(k_{F}x\left(1-F_{L}\frac{\exp\left(-2F_{L}/\pi c\right)}{\pi c}\right)\right)\cos\left(\frac{V}{2}x\left(1-F_{L}\frac{\exp\left(-2F_{L}/\pi c\right)}{\pi c}\right)\right)\label{eq:Final_field_field}
\end{align}

We now proceed to the density density calculation. We know that the
density density function is given by \cite{key-52-1}: 
\[
\left\langle \rho\left(x\right)\rho\left(0\right)\right\rangle =\rho^{2}-\frac{1}{4\pi^{2}}\int dk_{1}f_{L}\left(k_{1}\right)\omega\left(k_{1}\right)\int dk_{2}f_{L}\left(k_{2}\right)\omega\left(k_{2}\right)\left(\frac{k_{1}-k_{2}+ic}{k_{1}-k_{2}-ic}\right)\left(\frac{p\left(k_{1},k_{2}\right)}{k_{1}-k_{2}}\right)\exp\left(xp\left(k_{1},k_{2}\right)\right)
\]

Here $p\left(k_{1},k_{2}\right)=-i\left(k_{1}-k_{2}\right)+\int dtf_{L}\left(t\right)P_{t}\left(k_{1},k_{2}\right)$.
Here the function $P_{t}\left(k_{1},k_{2}\right)$ satisfies: 
\begin{equation}
2\pi P_{t}\left(k_{1},k_{2}\right)=\frac{k_{1}-t+ic}{k_{1}-t-ic}\cdot\frac{k_{2}-t-ic}{k_{2}-t+ic}\exp\left(-\int f_{L}\left(s\right)K\left(s,t\right)P_{s}\left(k_{1},k_{2}\right)\right)-1\label{eq:Genrating_function_density_density}
\end{equation}

From this we obtain that $P_{t}\left(k_{1},k_{2}\right)=\frac{-i}{\pi c}\left(k_{1}-k_{2}\right)+...$.
Combing we obtain that 
\begin{align*}
\left\langle \rho\left(x\right)\rho\left(0\right)\right\rangle \cong & \rho^{2}-\frac{1}{4\pi^{2}}\exp\left(-\frac{2F_{L}}{\pi c}\right)\left(1+\frac{F_{L}}{\pi c}\right)^{2}\int dk_{1}f_{L}\left(k_{1}\right)\int dk_{2}f_{L}\left(k_{2}\right)\left(1-2i\frac{k_{1}-k_{2}}{c}\right)\exp\left(-ix\left(k_{1}-k_{2}\right)\left(1+\frac{F_{L}}{\pi c}\right)\right)\\
= & \rho^{2}-\frac{4}{\pi^{2}}\exp\left(-\frac{2F_{L}}{\pi c}\right)\frac{1}{x^{2}}\sin^{2}\left(k_{F}x\left(1+\frac{F_{L}}{\pi c}\right)\right)\cos^{2}\left(\frac{V}{2}x\left(1+\frac{F_{L}}{\pi c}\right)\right)\\
 & -\frac{1}{\pi^{2}c}\exp\left(-\frac{2F_{L}}{\pi c}\right)\frac{1}{x^{3}\left(1+\frac{2F_{L}}{\pi c}\right)}\sin^{2}\left(k_{F}x\left(1+\frac{F_{L}}{\pi c}\right)\right)\cos^{2}\left(\frac{V}{2}x\left(1+\frac{F_{L}}{\pi c}\right)\right)\\
 & -\frac{1}{\pi^{2}c}\exp\left(-\frac{2F_{L}}{\pi c}\right)\frac{1}{x^{2}}\sin\left(k_{F}x\left(1+\frac{F_{L}}{\pi c}\right)\right)\cos\left(\frac{V}{2}x\left(1+\frac{F_{L}}{\pi c}\right)\right)\times\\
 & \times\left(V\left\{ \cos\left(x\left(\frac{V}{2}-k_{F}\right)\left(1+\frac{F_{L}}{\pi c}\right)\right)-\cos\left(x\left(\frac{V}{2}+k_{F}\right)\left(1+\frac{F_{L}}{\pi c}\right)\right)\right\} +\right.\\
 & +\left.2k_{F}\left\{ \cos\left(x\left(\frac{V}{2}-k_{F}\right)\left(1+\frac{F_{L}}{\pi c}\right)\right)+\cos\left(x\left(\frac{V}{2}+k_{F}\right)\left(1+\frac{F_{L}}{\pi c}\right)\right)\right\} \right)
\end{align*}

\end{widetext}As such to leading order in $1/c$ we have calculated
all the correlation functions for the two box quench.

\section*{Correlation Functions (BEC)}

We would like to carry out similar calculations to the ones done above
in the case when there are two boxes each of which is initialized
in a BEC each of length $l$ with $N$ particles. The boxes are moving
with velocities $V$ and $-V$ (the container box is assumed to have
size $L$). We will calculate the expectation values of the operators
$\left\langle b^{\dagger}\left(0\right)b^{\dagger}\left(0\right)b\left(0\right)b\left(0\right)\right\rangle $,
$\left\langle b^{\dagger}\left(0\right)b^{\dagger}\left(0\right)b^{\dagger}\left(0\right)b\left(0\right)b\left(0\right)b\left(0\right)\right\rangle $,
$\left\langle b^{\dagger}\left(x\right)b\left(0\right)\right\rangle $
and $\left\langle \rho\left(x\right)\rho\left(0\right)\right\rangle $.
We will work in the limit of large $c$ and to leading order in $1/c$.
We will also assume that both $x$ and $0$ are far away from the
box boundaries so that we may use the doubled box system to do all
calculations. The first step towards this calculation is to calculate
the occupation probability of the BEC quench $f\left(k\right)=\frac{\rho_{p}\left(k\right)}{\rho_{t}\left(k\right)}$.
It is known that for large $c$ the total quasiparticle density satisfies:
\begin{equation}
\rho_{t}\left(k\right)=\frac{1}{2\pi}+\frac{1}{\pi c}\int\rho_{p}\left(q\right)dq=\frac{1}{2\pi}+\frac{2N}{\pi cL}\label{eq:density}
\end{equation}

From this we obtain that 
\begin{equation}
f\left(k\right)=B_{L}\times\left(\rho^{BEC}\left(k-\frac{V}{2}\right)+\rho^{BEC}\left(k+\frac{V}{2}\right)\right)\label{eq:occupation}
\end{equation}

Here for future use we have defined $B_{L}=\frac{l}{L}\times\frac{1}{\frac{1}{2\pi}+\frac{2N}{\pi cL}}$.
Furthermore we note that for large $c$: $\rho^{BEC}\left(k\right)\simeq\frac{1}{2\pi}\frac{4n^{2}}{k^{2}+4n^{2}}+O\left(\frac{1}{c^{2}}\right)$
with $n=\frac{N}{L}$ \cite{key-44}. Next we know that \cite{key-49-1}:\begin{widetext}
\begin{align}
\left\langle b^{\dagger}\left(0\right)b^{\dagger}\left(0\right)b\left(0\right)b\left(0\right)\right\rangle  & =2\int\frac{dk_{1}}{2\pi}f_{L}\left(k_{1}\right)\int\frac{dk_{2}}{2\pi}f_{L}\left(k_{2}\right)\frac{\left(k_{1}-k_{2}\right)^{2}}{\left(k_{1}-k_{2}\right)^{2}+c^{2}}+....\nonumber \\
 & \cong\frac{1}{\pi^{2}}B_{L}^{2}N^{2}\left(\frac{V}{c}\right)^{2}+....\label{eq:density_density_zero-1}
\end{align}

Here we have assumed that $n\ll V\ll c$. Furthermore we may calculate
the density density density correlator similarly, it is given by \cite{key-49-1}:
\begin{align*}
\left\langle b^{\dagger}\left(0\right)b^{\dagger}\left(0\right)b^{\dagger}\left(0\right)b\left(0\right)b\left(0\right)b\left(0\right)\right\rangle  & \cong6\int\frac{dk_{1}}{2\pi}\int\frac{dk_{2}}{2\pi}\int\frac{dk_{3}}{2\pi}f_{L}\left(k_{1}\right)f_{L}\left(k_{2}\right)f_{L}\left(k_{3}\right)\frac{\left(k_{1}-k_{2}\right)^{2}}{\left(k_{1}-k_{2}\right)^{2}+c^{2}}\frac{\left(k_{1}-k_{3}\right)^{2}}{\left(k_{1}-k_{3}\right)^{2}+c^{2}}\frac{\left(k_{2}-k_{3}\right)^{2}}{\left(k_{2}-k_{3}\right)^{2}+c^{2}}\\
 & \cong\frac{9}{8\pi^{3}}B_{L}^{3}N\left(\frac{V}{c}\right)^{4}\times\frac{-\frac{c^{2}}{4n^{2}}-\frac{c}{2n}+\left(\frac{c^{2}}{4n^{2}}-1\right)\frac{c}{2n}+\frac{c^{2}}{4n^{2}}\sqrt{1+\frac{c^{2}}{4n^{2}}+\frac{c}{n}}-2\sqrt{\frac{c^{2}}{4n^{2}}\left(1+\frac{c^{2}}{4n^{2}}+\frac{c}{n}\right)}}{\left(\frac{c^{2}}{4n^{2}}-1\right)\sqrt{\frac{c^{2}}{4n^{2}}\left(1+\frac{c^{2}}{4n^{2}}+\frac{c}{n}\right)}}\\
 & \cong\frac{9}{4\pi^{3}}B_{L}^{3}N\left(\frac{V}{c}\right)^{4}\left(\frac{n}{c}\right)+...
\end{align*}

Here we have assumed that $n\ll V\ll c$. We can now calculate the
density density correlation function. We know that the density density
function is given by \cite{key-52-1}: 
\[
\left\langle \rho\left(x\right)\rho\left(0\right)\right\rangle =\rho^{2}-\frac{1}{4\pi^{2}}\int dk_{1}f_{L}\left(k_{1}\right)\omega\left(k_{1}\right)\int dk_{2}f_{L}\left(k_{2}\right)\omega\left(k_{2}\right)\left(\frac{k_{1}-k_{2}+ic}{k_{1}-k_{2}-ic}\right)\left(\frac{p\left(k_{1},k_{2}\right)}{k_{1}-k_{2}}\right)\exp\left(xp\left(k_{1},k_{2}\right)\right)
\]

Here $\omega\left(k\right)=\exp\left(-\frac{1}{2\pi}\int dqK\left(k,q\right)f\left(q\right)\right)\cong\exp\left(-\frac{G_{L}}{\pi c}\right)$
with $G_{L}=2B_{L}n$. Here $p\left(k_{1},k_{2}\right)=-i\left(k_{1}-k_{2}\right)+\int dtf_{L}\left(t\right)P_{t}\left(k_{1},k_{2}\right)$.
Here the function $P_{t}\left(k_{1},k_{2}\right)$ satisfies: 
\begin{equation}
2\pi P_{t}\left(k_{1},k_{2}\right)=\frac{k_{1}-t+ic}{k_{1}-t-ic}\cdot\frac{k_{2}-t-ic}{k_{2}-t+ic}\exp\left(-\int f_{L}\left(s\right)K\left(s,t\right)P_{s}\left(k_{1},k_{2}\right)\right)-1\label{eq:Genrating_function_density_density-1}
\end{equation}

From this we obtain that $P_{t}\left(k_{1},k_{2}\right)=\frac{-i}{\pi c}\left(k_{1}-k_{2}\right)+...$.
Furthermore $\frac{k_{1}-k_{2}+ic}{k_{1}-k_{2}-ic}\simeq-1\left(1-2\frac{i}{c}\left(k_{1}-k_{2}\right)\right)\simeq-\exp\left(-2\frac{i}{c}\left(k_{1}-k_{2}\right)\right)$.
We now obtain that $p\left(k_{1},k_{2}\right)\cong-i\left(k_{1}-k_{2}\right)\left(1+\frac{G_{L}}{\pi c}\right)$.
Combining we obtain that: 
\begin{align}
\left\langle \rho\left(x\right)\rho\left(0\right)\right\rangle  & =\rho^{2}-\frac{2+2\cos\left(\frac{V}{2}\left(x\left(1+\frac{G_{L}}{\pi c}\right)+\frac{2}{c}\right)\right)}{4\pi^{2}}\exp\left(-\frac{G_{L}}{\pi c}\right)\left(1+\frac{G_{L}}{\pi c}\right)^{2}B_{L}^{2}\times\nonumber \\
 & \times\int dk_{1}\rho^{BEC}\left(k_{1}\right)\int dk_{2}\rho^{BEC}\left(k_{2}\right)\exp\left(-i\left(k_{1}-k_{2}\right)\left(x\left(1+\frac{G_{L}}{\pi c}\right)+\frac{2}{c}\right)\right)=\nonumber \\
 & \rho^{2}-\frac{2+2\cos\left(\frac{V}{2}\left(x\left(1+\frac{G_{L}}{\pi c}\right)+\frac{2}{c}\right)\right)}{4\pi^{2}}\exp\left(-\frac{G_{L}}{\pi c}\right)\left(1+\frac{G_{L}}{\pi c}\right)^{2}B_{L}^{2}\cdot n^{2}\exp\left(-2n\left(x\left(1+\frac{G_{L}}{\pi c}\right)+\frac{2}{c}\right)\right)\label{eq:Density_density}
\end{align}

We would now like to calculate the field-field correlation function.
It is given by \cite{key-53}: 
\begin{equation}
\left\langle b^{\dagger}\left(x\right)b\left(0\right)\right\rangle \cong\int\frac{dk}{2\pi}f_{L}\left(k\right)e^{-ikx}\omega\left(k\right)\times\exp\left(-x\int dtf_{L}\left(t\right)P_{t}\left(k\right)\right)\label{eq:Field_Field}
\end{equation}

Here $\omega\left(k\right)=\exp\left(-\frac{1}{2\pi}\int dqK\left(k,q\right)f\left(q\right)\right)\cong\exp\left(-\frac{G_{L}}{\pi c}\right)$,
where and $K\left(k,q\right)=\frac{2c}{\left(k-q\right)^{2}+c^{2}}\cong\frac{2}{c}$.
Furthermore the function $P_{t}\left(k\right)$ satisfied the equation:
\begin{equation}
2\pi P_{t}\left(k\right)=-\frac{k-t+ic}{t-k+ic}\exp\left(-\int f_{L}\left(s\right)K\left(t,s\right)P_{s}\left(k\right)\right)-1\label{eq:Field_Field_generating_function-1}
\end{equation}

Using this expression it is possible to obtain that: 
\[
P_{t}\left(k\right)\cong-\frac{1}{2\pi}\left(1+\exp\left(-\frac{2G_{L}}{\pi c}\right)\right)+i\frac{\exp\left(-\frac{2G_{L}}{\pi c}\right)}{\pi c}\left(k-t\right)
\]

From this we obtain that $\int f_{L}\left(t\right)P_{t}\left(k\right)\cong-\frac{G_{L}}{2\pi}\left(1+\exp\left(-\frac{2G_{L}}{\pi c}\right)\right)+iG_{L}\frac{k}{\pi c}\exp\left(-\frac{2G_{L}}{\pi c}\right)$.
Combing we obtain that : 
\begin{align*}
\left\langle b^{\dagger}\left(x\right)b\left(0\right)\right\rangle  & =\frac{\exp\left(-\frac{G_{L}}{\pi c}\right)}{2\pi}\exp\left(-\frac{G_{L}x}{2\pi}\left(1+\exp\left(-\frac{2G_{L}}{\pi c}\right)\right)\right)\int f_{L}\left(k\right)e^{-ikx\left(1-G_{L}\frac{\exp\left(-2G_{L}/\pi c\right)}{\pi c}\right)}\\
 & =\frac{\exp\left(-\frac{G_{L}}{\pi c}\right)}{\pi}\exp\left(-\frac{G_{L}x}{2\pi}\left(1+\exp\left(-\frac{2G_{L}}{\pi c}\right)\right)\right)B_{L}\cos\left(\frac{V}{2}\left(1-G_{L}\frac{\exp\left(-2G_{L}/\pi c\right)}{\pi c}\right)\right)\times\\
 & \times\int\rho^{BEC}\left(k\right)e^{-ikx\left(1-G_{L}\frac{\exp\left(-2G_{L}/\pi c\right)}{\pi c}\right)}\\
 & =\frac{\exp\left(-\frac{G_{L}}{\pi c}\right)}{\pi}\exp\left(-\frac{G_{L}x}{2\pi}\left(1+\exp\left(-\frac{2G_{L}}{\pi c}\right)\right)\right)B_{L}\times\\
 & \times\cos\left(\frac{V}{2}\left(1-G_{L}\frac{\exp\left(-2G_{L}/\pi c\right)}{\pi c}\right)\right)n\exp\left(-2nx\left(1-G_{L}\frac{\exp\left(-2G_{L}/\pi c\right)}{\pi c}\right)\right)
\end{align*}

The velocity probably distribution is then: 
\begin{equation}
P\left(v\right)\sim\int dxe^{-i\frac{v}{2}x}\left\langle b^{\dagger}\left(x\right)b\left(0\right)\right\rangle \sim nB_{L}\frac{\exp\left(-\frac{G_{L}}{\pi c}\right)}{\pi}\times\left(\frac{H_{L}}{H_{L}^{2}+\frac{1}{4}\left(v-VK_{L}\right)}+\frac{H_{L}}{H_{L}^{2}+\frac{1}{4}\left(v+VK_{L}\right)}\right)\label{eq:probality_BEC}
\end{equation}

Here $H_{L}=\frac{G_{L}}{2\pi}\left(1+\exp\left(-\frac{2G_{L}}{\pi c}\right)\right)+2n\left(1-G_{L}\frac{\exp\left(-2G_{L}/\pi c\right)}{\pi c}\right)$
and $K_{L}=\left(1-G_{L}\frac{\exp\left(-2G_{L}/\pi c\right)}{\pi c}\right)$.

\end{widetext}

\section*{q-boson regularization }

We wish to show that the in the thermodynamic limit the edge contributions
to the conserved quantities vanish. To do so we need to introduce
a q-boson regularization of the conserved charges \cite{key-45,key-47}.
The q-boson system corresponds to $M$ bosonic lattice sites with
each site having operators $B_{n}$, $B_{n}^{\dagger}$ and $N_{n}=N_{n}^{\dagger}$
that satisfy the relations $B_{n}B_{n}^{\dagger}-q^{-2}B_{n}^{\dagger}B_{n}=1$,
$\left[N_{n},B_{n}\right]=-B_{n}$ and $\left[N_{n},B_{n}^{\dagger}\right]=B_{n}^{\dagger}$.
The q-boson Hamiltonian is given by: 
\begin{equation}
H_{q}=-\frac{1}{\delta^{2}}\sum_{n=1}^{M}\left(B_{n}^{\dagger}B_{n+1}+B_{n+1}^{\dagger}B_{n}-2N_{n}\right)\label{eq:q-boson_Hamiltonian}
\end{equation}

The system is integrable since the Hamiltonian may be derived from
the following transfer matrix 
\begin{equation}
T=\left(\begin{array}{cc}
A\left(\lambda\right) & B\left(\lambda\right)\\
C\left(\lambda\right) & D\left(\lambda\right)
\end{array}\right)=L_{M}\left(\lambda\right)....L_{1}\left(\lambda\right)\label{eq:Transfer_matrix}
\end{equation}

With 
\begin{equation}
L_{n}\left(\lambda\right)=\left(\begin{array}{cc}
e^{\lambda} & \chi B_{n}^{\dagger}\\
\chi B_{n} & e^{-\lambda}
\end{array}\right)\label{eq:L-Operator}
\end{equation}

Here $\chi=\sqrt{1-q^{-2}}$ and $q=e^{\lambda}$. There is an infinite
family of conserved charges $I_{n}$, the first few densities corresponding
to these conserved charges are given by: 
\begin{align}
J^{1}\left(n\right)= & \chi^{2}B_{n}^{\dagger}B_{n+1}\nonumber \\
J^{2}\left(n\right)= & \chi^{2}\left(1-\frac{\chi^{2}}{2}\right)\left(B_{n}^{\dagger}B_{n+2}-\right.\nonumber \\
 & \left.-\frac{\chi^{2}}{2-\chi^{2}}B_{n}^{\dagger}B_{n}^{\dagger}B_{n+1}B_{n+1}-\chi^{2}B_{n}^{\dagger}B_{n+1}^{\dagger}B_{n+1}B_{n+2}\right)\label{eq:Conserved_charges}
\end{align}

Furthermore the open q-boson chain is also integrable \cite{key-48}.
It is known that the Lieb-Liniger gas is a limiting case of the q-bosons,
where the limit is taken as $\delta\rightarrow0$, $M\delta=L$, $\gamma=\frac{c\delta}{2}$
and $q=e^{\lambda}$. We shall show that in the limit $L_{i}\rightarrow\infty$
for any finite $\chi$ the edges give no contribution to the conserved
quantities. Indeed we notice that the conserved quantities are linear
functions of the expectations of various operators e.g. $I^{1}=\sum_{n}\left\langle B_{n}^{\dagger}B_{n+1}\right\rangle $
with $\sim L_{i}$ terms in the sum. Furthermore by translational
invariance each of the terms gives the same contribution e.g. 
\begin{equation}
I^{i}=\frac{L_{i}}{\delta}\left\langle J^{i}\left(n_{0}\right)\right\rangle -Boundary\, Terms\label{eq:Conserved_charges-1}
\end{equation}

Here $n_{0}$ is some site in the middle of the q-boson chain. We
notice that the expectation values of the boundary terms have absolutely
no $L$ dependence (they are just proportional to the expectation
value of the density, density density, field-field and related correlation
functions which do not scale with $L$). Therefore in the limit that
$L_{i}\rightarrow\infty$ we have that $I^{i}=\frac{L_{i}}{\delta}\left\langle J^{i}\left(n_{0}\right)\right\rangle $
and the boundary terms have disappeared. Since the Lieb-Liniger gas
corresponds to a limit of the q-bosons we see that it the thermodynamic
limit the boundary terms do not effect conserved quantities.

\section*{Initial Correlations}

We would like to calculate the velocity probability distribution when
the traps are initially released at time equal to zero. This would
help us compare with the time averaged case. The experimentally accessible
quantities are most easily given in terms of an average velocity probability:
\begin{equation}
P_{av}\left(v\right)=\frac{1}{L}\int dx\int dye^{-i\frac{v}{2}x}\left\langle b^{\dagger}\left(x\right)b\left(y\right)\right\rangle \label{eq:Average_correlation}
\end{equation}

In the case of the BEC it is not too hard to see that 
\begin{equation}
P_{av}\left(v\right)=\frac{l}{L}n\left(\delta\left(v-V\right)+\delta\left(v+V\right)\right)\label{eq:Average_Prob}
\end{equation}

In the case of the two boxes in their ground state, following a derivation
given above we see that the velocity probably distribution:

\begin{align}
P_{av}\left(v\right)\sim\frac{l}{L}\frac{\exp\left(-\frac{J_{L}}{\pi c}\right)}{2\pi}\sum_{i,j=\pm}\left(-1\right)^{j}\arctan A_{i,j}(v)\label{velProb-1}
\end{align}
with $A_{\pm\pm}(v)=C_{L}\left((1-J_{L}\frac{\exp\left(-2J_{L}/\pi c\right)}{\pi c})\left(\pm\frac{V}{2}\pm k_{F}\right)+\frac{v}{2}\right)$,
and $C_{L}=\frac{2\pi}{J_{L}\left(1+\exp\left(-\frac{2J_{L}}{\pi c}\right)\right)}$,
with $J_{L}=2k_{F}$. These results are used in Fig. \ref{fig:Poles}(B-G). 

\begin{thebibliography}{10}
\bibitem[45]{key-54} We note that these conserved quantities may
need to be regularized \cite{key-46,key-45}.
\end{thebibliography}
\end{document}